 \documentclass{emulateapj}

\usepackage{amsmath}
\slugcomment{Draft Version \today}
\shorttitle{Tidal Stream Morphology as an Indicator of Dark Matter Halo Geometry}
\shortauthors{PEARSON ET AL.}
\usepackage[backref,breaklinks,colorlinks,citecolor=blue]{hyperref}
\usepackage[all]{hypcap}

\begin{document}

\title{Tidal Stream Morphology as an Indicator of Dark Matter Halo Geometry:\\ the Case of Palomar\,5}

\author{Sarah Pearson\altaffilmark{1}, Andreas H. W. K\"{u}pper\altaffilmark{1,}\altaffilmark{2}, Kathryn V. Johnston\altaffilmark{1},  Adrian M. Price-Whelan\altaffilmark{1}}
\altaffiltext{1}{Department of Astronomy, Columbia University, New York, NY 10027, USA}
\altaffiltext{2}{Hubble Fellow}

\email{spearson@astro.columbia.edu}

\begin{abstract}
This paper presents an example where the morphology of a single stellar stream can be used to rule out a specific galactic potential form without the need for velocity information.
We investigate the globular cluster Palomar\,5 (Pal\,5), which is tidally disrupting into a cold, thin stream mapped over 22 degrees on the sky with a typical width of 0.7 degrees. 
We generate models of this stream by fixing Pal\,5's present-day position, distance and radial velocity via observations, while allowing its proper motion to vary. In a spherical dark matter halo we easily find models that fit the observed morphology. However, no plausible Pal\,5 model could be found in the triaxial potential of \citet{lawmajew10}, which has been proposed to explain the properties of the Sagittarius stream. In this case, the long, thin and curved morphology of the Pal\,5 stream alone can be used to rule out such a potential configuration. Pal\,5 like streams in this potential are either too straight, missing the curvature of the observations, or show an unusual morphology which we dub \textit{stream-fanning}:  a signature sensitive to the triaxiality of a potential.
We conclude that the mere existence of other thin tidal streams must provide broad constraints on the orientation and shape of the dark matter halo they inhabit.
\end{abstract}


\section{Introduction} \label{sec:intro}
Simulations of large-scale structure formation suggest that all galaxies lie within triaxial dark matter halos (e.g., \citealt{bullock02}). The Milky Way (MW) offers a unique perspective on this problem as it is the one galaxy not seen in projection and therefore the one galaxy where we can measure the true 3D shape and orientation of a dark matter halo. However, constraints from observations on the shape of the Milky Way's dark halo remain uncertain and inconsistent. 

The distances, radial velocities, and positions of stars in the Sagittarius stream provide a rich data set to use to probe our dark matter halo. \citet{lawmajew10} (LM10) were the first to attempt modeling the MW including a fully triaxial dark matter halo using the data that were then available. Their best-fit halo model predicts an almost oblate dark matter configuration oriented perpendicular to the Galactic disk. However, concerns have been raised questioning the validity of the LM10 potential: i) \citet{Deb13} showed that the orientation of the halo in this model could not host a stable disk; ii) \citet{Ibata13} suggested that it is possible to approximately reproduce the spatial and kinematic structure of the Sagittarius stream without introducing triaxiality to the dark matter halo component (although their paper does not include a quantitative assessment that their model fitted the data as well as the LM10 simulation); and iii) \citet{belokurov14} demonstrated that the LM10 model fits neither the extent  nor the precession angle between successive apocenters in a newly found part of the Sagittarius stream. A solution to the first part of the problem was proposed by \citet{cirohelmi13}: by introducing a transition from an oblate halo at small radii to a triaxial LM10 halo on larger scales, a potential could be obtained that could host both the Sagittarius stream and the Galactic disk. However, there is not yet a model that successfully reproduces the full data set (i.e. including the more recent evidence for Sagittarius material extending out to Galactocentric distances of more than 100 kpc found by \citet{belokurov14}), in its entirety.\footnote{See \citealt{gib14} for an interesting discussion of what the distances to and angular positions of the apocenters {\it alone} might tell us about the radial density profile of the dark matter halo.} For this reason it is important to test any suggested potential form with other streams than just the Sagittarius stream. 

In this article, we look at the tidal stream originating from Palomar\,5 (Pal\,5), a globular cluster currently at the apocenter of its orbit, 23.6 kpc from the Sun \citep{dotter11}. Pal\,5 is orbiting the MW at a much smaller radius than Sagittarius and thus serves as a probe of the shape of the MW in a different region of the Galaxy. Pal\,5's tidal stream was first discovered by \citet{oden01} and was subsequently mapped over 22 degrees on the sky \citep{grill06} with a typical width of 0.7 degrees \citep{oden03,carlberg12}.

In what follows, we explore what the morphology of Pal\,5's tidal tails can tell us about the gravitational potential of the MW by using a combination of restricted three-body models \citep{kupper12} and $N$-body simulations. Our model streams are simulated in two different Milky-Way-like potentials consisting of a disk and bulge embedded within a dark matter halo that is either spherical or triaxial in shape. We show that in a LM10-like halo configuration, our simulated Pal5 streams cannot match the thin, ``S''-shape and curved morphology of the observed stellar density: they are either too straight or exhibit a broad morphology which we dub \textit{stream-fanning}. Thus the thin and curved morphology of Pal 5 alone has given us a fast and simple way to check if this particular potential form is realistic. The broader implication of this simple test is that the mere existence of many such thin streams at different distances and orientations around our Galaxy can rule out other classes of triaxial potentials.

In Section \ref{sec:methods}, we describe the methods used to simulate the morphology of Pal\,5's tidal tails. In Section \ref{sec:results} we compare the streams produced in \textit{streakline} models and $N$-body simulations to observed data within the two test-potentials. We first do a comparison to over-densities of Pal\,5 stars from SDSS only, then include radial velocities of Pal\,5 stars in the fit. In Section \ref{sec:discussion}, we investigate further possible parameter variations, and we discuss possible origins of \textit{stream-fanning} in Section \ref{sec:morph}. We conclude in Section \ref{sec:conclusion}.

\section{Methods} \label{sec:methods}
In this work we compare two different trial galactic potentials having either a spherical or triaxial dark matter halo, which we introduce in Section \ref{sec:pot}. In Section \ref{sec:streakline} and \ref{sec:obs}, we describe how the most likely orbit within a given potential is found through a comparison of \textit{streakline} model streams with observational data. We then describe the $N$-body simulations used to illustrate  our results in Section \ref{sec:Nbody}. 

\subsection{Form of the Galactic potential}\label{sec:pot}
In our \textit{streakline} and $N$-body simulations, the potential of the MW is computed consisting of a disk and bulge embedded within a dark matter halo that is either spherical or triaxial in shape. We approximate the baryonic component of the MW using a \citet{miya75} disk ($M_{disk} = 10^{11} \mathrm{M}_{\odot}$, $a$ = 6.5 kpc, $b$ = 0.26 kpc), and a Hernquist spheroid for the bulge ($M_{bulge}$ = 3.4 $\times$ 10$^{10} \mathrm{M}_{\odot}$ and  $c$ = 0.7 kpc) (\citealt{hern90}). This parametrization of the Galactic disk and bulge was chosen for computational simplicity and is used widely in the literature. More realistic forms of disk and bulge have been proposed by, e.g., \citet{dehnen98}, however, our focus lies on the comparison of our models to the previous work of LM10. For the dark matter component we use two different halo potential forms: 
\begin{enumerate}
\item \textit{Triaxial dark matter halo}: following LM10, we parametrize the halo potential as
\begin{eqnarray}\label{lm10}
\Phi_{halo} &=& v^2_{halo} \mbox{ln}(C_1x^2 + C_2y^2 + C_3xy + \frac{z^2}{q_z^2} + r_{halo}^2)\\
C_{1} &=& \left(\frac{\cos^2\phi}{q_1^2} + \frac{\sin^2\phi}{q_2^2}\right)\\
C_{2} &=& \left(\frac{\cos^2\phi}{q_2^2} + \frac{\sin^2\phi}{q_1^2}\right)\\
C_{3} &=& 2 \sin \phi \cos\phi\left(\frac{1}{q_1^2} - \frac{1}{q_2^2}\right)
\end{eqnarray}
where we use the exact same parameters for the triaxial dark matter halo as LM10. That is, the rotation angle of the $x$-axis around $z$ from the Sun-Galactic center line is $\phi = 97$ deg, the ratios between where the equipotential contours intersect the $x/y$ and $z/y$ axes are $q_1$= 1.38, $q_z$ = 1.36 respectively. $q_2$ = 1.0 by definition, $v_{halo}$ = 121.9 km/s and $r_{halo}$ = 12.0 kpc. 

\item \textit{Spherical dark matter halo}: we use the same potential form as above for the spherical halo potential (Equation \ref{lm10}), but now $q_1$= 1.0, $q_2$= 1.0  and $q_z$ = 1.0. We set $v_{halo}$ = 172.3 km/s and $r_{halo}$ = 12.0 kpc to ensure $v_{rot}$ = 220 km/s at $R$ = 8.3 kpc.  
\end{enumerate}
The rotation curves of these two galactic potentials match the overall shape of observed MW rotation curves (cf.~\citealt{sofue13, Irrgang14}). In this configuration, the Sun sits at $\vec{R}_{\odot} = (-8.3, 0, 0)$\,kpc, with a velocity of $\vec{V}_{\odot} = (11.1, 258.1, 7.3)$\,km/s (\citealt{gillessen09}, \citealt{schonrich12}, \citealt{reid14}, \citealt{kuepper15}).

\begin{figure}
\centerline{\includegraphics[width=\columnwidth]{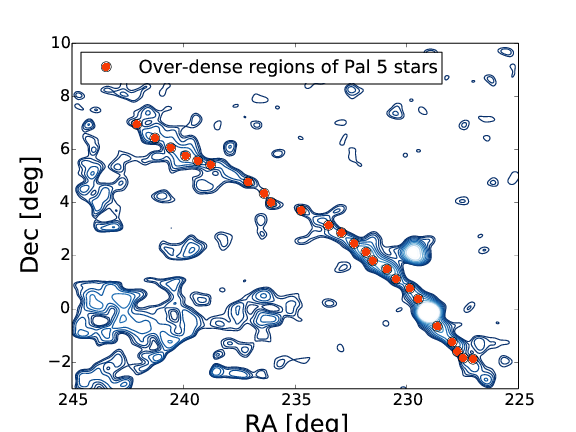}}
\caption{Matched-filter map of Pal\,5-like stars from SDSS DR9 (blue contours). We used the over-dense regions marked as orange points to assess the likelihood of our Pal\,5 \textit{streakline} models (\citealt{balb11}, \citealt{kuepper15}).}
\label{fig:OD}
\end{figure}
\subsection{The \textit{Streakline} method}\label{sec:streakline}
To create model streams along a given orbit in a specific potential, we use the \textit{streakline} method outlined in \citet{kupper12}, \citet{lane12}, \citet{bonaca14} and \citet{kuepper15}, which is closely related to the methods used in \citet{varghese11} and \citet{gib14}. \citet{bonaca14} demonstrated that the \textit{streakline} method is a computationally efficient way of generating realistic streams that match the morphology of much more time-consuming full $N$-body models. \textit{Streakline} models are restricted three-body models of tidal streams: the dissolution of a star cluster due to the tidal field of its host galaxy is approximated by a ``star-cluster particle'' orbiting within an analytic galaxy potential that releases test particles at a given time interval. The test particles are then integrated together with the cluster particle within the background potential. The test particles do not interact with each other, which makes the \textit{streakline} method very fast. However, the gravitational attraction of the cluster particle on the released test particles is included, which was shown to be of importance for reproducing the morphology and length of streams from full $N$-body models \citep{kupper12, gib14}. For simplicity, the cluster particle is represented by a smoothed point mass with a smoothing length of 20 pc, which is of the order of Pal\,5's half-light radius \citep{oden03}. 

Our \textit{streakline} model assumes that stars escape from the cluster at the tidal radius \citep{king62},
\begin{equation}
r_t= \left(\frac{GM(t)}{\Omega^2 - \partial^2\Phi/\partial R^2}\right)^\frac{1}{3},
\end{equation}
at a constant rate. Here, $M(t)$ is the mass of the cluster at time $t$ (the final, i.e.~present-day, mass is $M$ = 15000$\mathrm{M}_{\odot}$; \citet{kuepper15}), $\Omega$ is the instantaneous angular velocity of the cluster with respect to the galactic center, $\Phi$ is the galactic potential, and $R$ is the cluster's current galactocentric distance. To introduce some scatter into these idealized escape conditions, we apply a random Gaussian spatial offset with a width of 0.25 $\times$ $r_t$ around the Lagrangian points at the time of escape \citep{lane12, bonaca14, gib14}. The stars are given velocities matching the angular velocity of the cluster plus an additional random Gaussian velocity offset with a dispersion of 1 km/s, comparable to the velocity dispersion of the cluster. Furthermore, we assume that the cluster itself has a constant mass loss rate of 8 M$_{\odot}$ per Myr, which was chosen based on $N$-body simulations of the cluster \citep{kuepper15}. However, we found that changing this mass loss rate has little to no effect on the results. 

From the six phase-space coordinates that determine Pal\,5's orbit, we fix the sky position (RA = 229.0$^{\circ}$, Dec =  -0.111$^{\circ}$), the radial velocity ($v_r$ = -58.7 km/s; \citealt{oden02}), and the distance\footnote{We explore the effect of varying the distance in Section \ref{sec:discussion}.} ($d$ = 23.6 kpc; \citealt{dotter11}). Hence, the only free parameters are the two proper motion components, which we vary in order to find the most likely orbit in each potential. For a given choice of phase-space coordinates, we integrate the orbit backwards for 6 Gyr and subsequently integrate it forward again while producing the \textit{streakline} model. After releasing test particles uniformly in time from the Lagrange points and integrating them to the present day, we compare the test particle distribution to the observations and assess the likelihood of the respective model.

\subsection{Comparison to observational data}\label{sec:obs}
For a given orbit within a specific potential, we compare the \textit{streakline} models to 24 over-dense regions of color-selected Pal\,5 stream stars shown in Figure \ref{fig:OD} (\citealt{balb11}, \citealt{kuepper15}). These regions were found through a \textit{Difference-of-Gaussians} process, in which a matched-filter map of Pal\,5 from SDSS data is smoothed with a small and a large Gaussian kernel, and the two maps are subtracted from each other. On the residuals map an extended-source finder algorithm like \textsc{SExtractor} \citep{Bertin96} is run (see \citealt{Koposov08} and \citealt{kuepper15} for details). The locations shown in Figure \ref{fig:OD} give the barycenters of extended, over-dense regions, which stood out by at least 3$\sigma$ above the random fluctuations in the residuals map. These over-densities give the regions with the highest local probability of finding Pal\,5 stars, and hence we require our models to go through these points.

Our comparison of the models to the observed over-densities is based on the framework developed in \citet{Hogg10} and \citet{Hogg12}, and applied to Pal\,5 in \citet{kuepper15}. We use a likelihood of the form:
\begin{equation}
\mathcal{L}_\mathrm{OD} = \prod^{N_{OD}}\frac{1}{{N}_\mathrm{model}}\sum_i^{{N}_{\mathrm{model}}} \left( \frac{1}{\sqrt{2\pi \Delta d^2}} \exp\left[-\frac{1}{2}\left(\frac{d_{ij}^2}{\Delta d^2}\right)\right] + \Delta \right)
\label{eq:LogL}
\end{equation}
Here $N_{\mathrm{OD}}$ is the number of over-densities, $d_{ij}$ is the distance from each model point to the $j$-th over-density, $\Delta d$ is the uncertainties in the barycenter positions of the over-densities, determined from the extended-source finding algorithm (\textsc{SExtractor}) and $\Delta$ is a numerical constant set to $\Delta = 10^{-5}$. This constant allows each over-dense region to not be a part of the stream by limiting its contribution to the likelihood to a minimum value. The maximum likelihood \emph{streakline} model maximizes the density of model points around the observed over-densities from SDSS.

Here we do not assume or imply anything about the origin of these over-densities. However, it has been shown that over-densities in tidal tails can be produced by epicyclic motion of stars evaporating from star clusters (\citealt{kupper10}). This may indeed be the origin of some of the over-densities closer to the cluster itself; for example, \citet{Mastrobuono12} have shown that tidal tails of Pal\,5-like clusters in fact show significant over-densities. Other possible explanations for inhomogeneities in tidal streams are variations in the mass loss rate, perturbations by dark matter subhalos, or variations in the depth of the observed data (e.g., \citealt{ngan14}). However, the primary purpose of our likelihood function is to assess the alignment of our generated models with the observed streams. An alternative approach would be to instead measure the smallest distance of each point from the centroid of the stream, but this effectively assumes that the density along the stream is constant. For a detailed analysis of the differences between these two different methods see \citet{kuepper15}.

We also make use of kinematic data from the literature. \citet{oden09} measured 17 radial velocities of stars in Pal\,5's tidal streams. When these are included in the assessment of the likelihoods, the full likelihood is: 
\begin{align}
\mathcal{L} &= \mathcal{L}_\mathrm{OD} \times \mathcal{L}_{v_r}\\
\ln \mathcal{L} &= \ln \mathcal{L}_\mathrm{OD} + \ln \mathcal{L}_{v_r} \label{eq:LogLtot}
\end{align}
where $\ln \mathcal{L}$ is the log-likelihood (LL) and $\mathcal{L}_{v_r}$ has the same form as Equation~\ref{eq:LogL}, but includes a comparison between the radial velocities of our models with the 17 radial velocities observed for Pal\,5. 

We have fixed all potential parameters in both potentials,  and can thus compare the likelihoods between the two potential forms by first using the over-densities only (Equation \ref{eq:LogL}) and then using both over-densities and radial velocities (Equation \ref{eq:LogLtot}).

\begin{figure*}
\centerline{\includegraphics[width=1.17\textwidth]{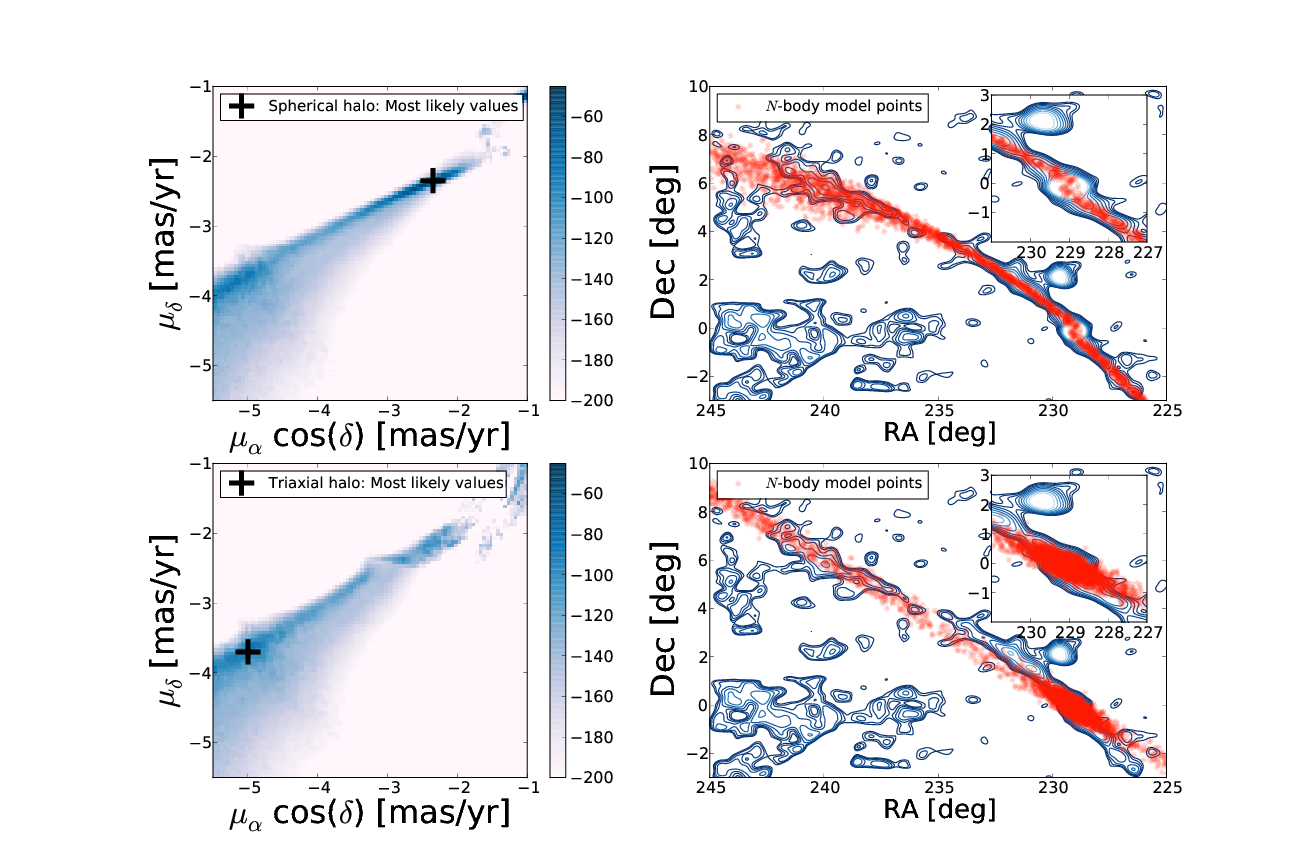}}
\caption{Left panels: log-likelihood value (color bar) of various proper motion configurations in the spherical potential (top) and triaxial LM10 potential (bottom) computed from \textit{streakline} models using Equation \ref{eq:LogL}. Right panels: \textsc{Nbody6} model points (orange) of the the most likely proper motion configuration in the spherical potential (top: ($\mu_{\delta}$, $\mu_{\alpha}\mathrm{cos}(\delta)$) = (-2.35, -2.35) mas yr$^{-1}$) and triaxial LM10 potential (bottom: ($\mu_{\delta}$, $\mu_{\alpha}\mathrm{cos}(\delta)$) = ( -3.7,- 5.0) mas yr$^{-1}$), over-plotted on SDSS density contours (blue). The \textit{streakline} model in the triaxial LM10 potential (LL = -82) yields a much lower log-likelihood, than the spherical case (LL = -45).}
\label{fig:posonly}
\end{figure*}

\subsection{$N$-body simulations} \label{sec:Nbody}
We construct $N$-body model streams of Pal\,5 by using the collisional $N$-body code \textsc{Nbody6} (\citealt{aarseth99, aarseth03}). \textsc{Nbody6} is GPU-enabled, which allows us to compute realistic Pal\,5 model streams on a star-by-star basis over several Gyr within a day (\citealt{nitaaarseth12}). 

To set up the initial cluster conditions for Pal\,5, we use the publicly available code \textsc{McLuster}\footnote{\url{https://github.com/ahwkuepper/mcluster}} (\citealt{kupper11b}). The initial number of stars is set to $N = 65536$ with stellar masses of 0.4\,M$_\odot$, following a Plummer density profile (\citealt{plummer1911}). We fix the radial velocity, present-day sky position and present-day distance of Pal\,5 to the observationally constrained values specified in Section \ref{sec:streakline}. We determine the two proper motion components of the cluster by exploring the \textit{streakline} model streams described in Section \ref{sec:streakline} and \ref{sec:obs}. For each setup, we ran a number of $N$-body models with initial half-mass radii in the range 10-20 pc for 6 Gyr, and picked the model with a final cluster mass close to Pal\,5's present-day mass of about 15,000\,M$_\odot$ (\citealt{kuepper15}). These models will be discussed in the following Section.

\begin{figure*}
\centerline{\includegraphics[width=1.17\textwidth]{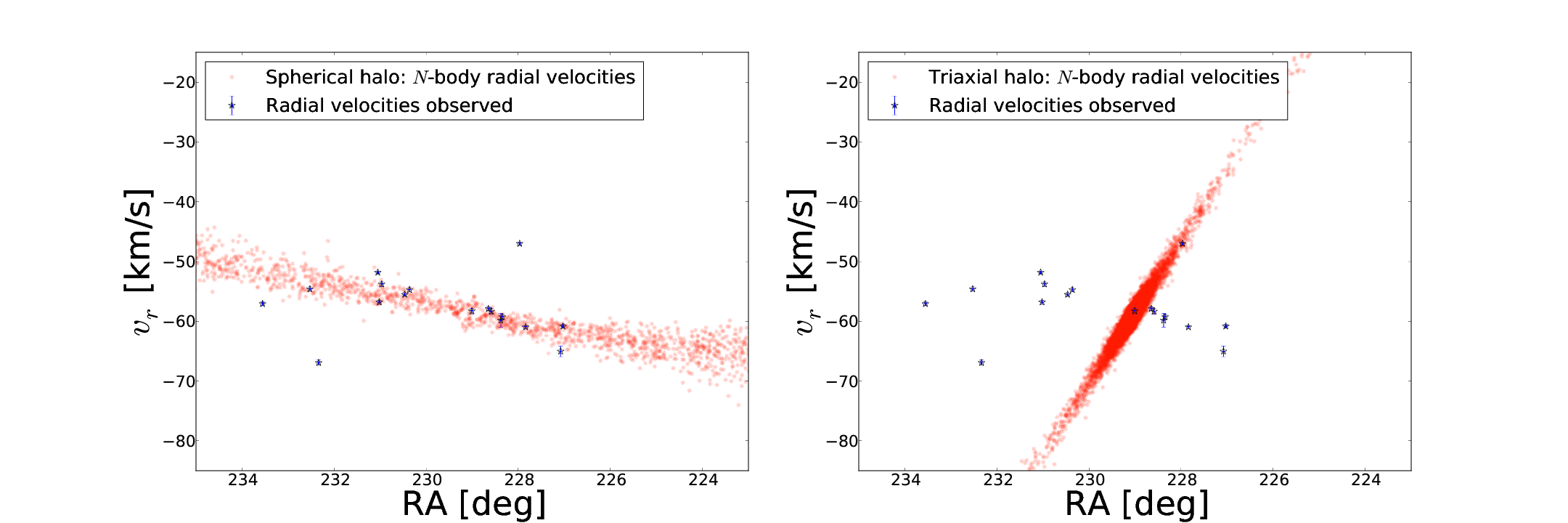}}
\caption{Line-of-sight velocities of \textsc{Nbody6} model points (orange) along the stream from the most likely proper motion configuration in the spherical potential (left: ($\mu_{\delta}$, $\mu_{\alpha}\mathrm{cos}(\delta)$) = (-2.35 , -2.35) mas yr$^{-1}$) and the triaxial LM10 potential (right: $(\mu_{\delta}$, $\mu_{\alpha}\mathrm{cos}(\delta)$  = ( -3.7, -5.0) mas yr$^{-1}$), plotted with the observed line-of-sight velocities (blue) from \citet{oden09}.}
\label{fig:radvel}
\end{figure*}

\begin{figure*}
\centerline{\includegraphics[width=1.17\textwidth]{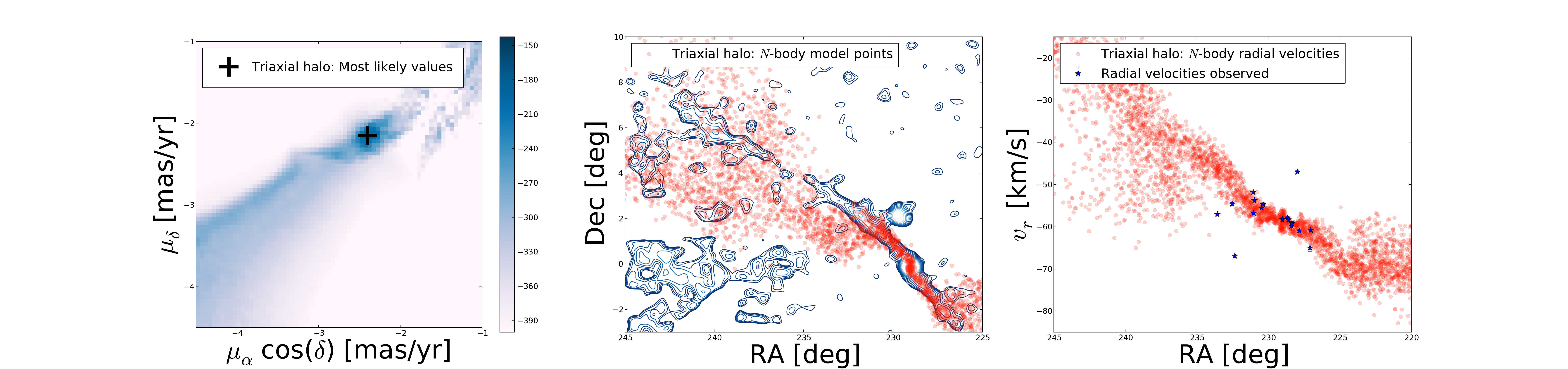}}
\caption{\textbf{Left panel}: log-likelihood value (color bar) of various proper motion configurations in the triaxial LM10 potential computed from \textit{streakline} models with an integration time of 6 Gyr. Distance, radial velocity and position were fixed. The log-likelihoods are calculated using Equation \ref{eq:LogLtot}. \textbf{Middle panel}: \textsc{Nbody6} model points (orange) of the the most likely proper motion configuration (($\mu_{\delta}$, $\mu_{\alpha}\mathrm{cos}(\delta)$) = (-2.15, -2.4) mas yr$^{-1}$), over-plotted on SDSS density contours (blue). \textbf{Right panel}: Line-of-sight velocities of \textsc{Nbody6} model points (orange), plotted with the observed line-of-sight velocities (blue) from \citealt{oden09}. The line-of-sight velocities of the $N$-body model points trace the observed gradient.}
\label{fig:posvr}
\end{figure*}

\section{Results} \label{sec:results}
Using the procedure outlined in Section \ref{sec:methods}, we examine the morphology of Pal\,5 in the spherical and the triaxial LM10 halo potentials. We first run \textit{streakline} models over a grid of reasonable proper motions in each potential for 6 Gyr, where we assess our likelihoods by comparing to over-densities in Pal\,5 only (Equation \ref{eq:LogL}). The results from this analysis are shown in the two left panels of Figure \ref{fig:posonly}. 

We find that the maximum likelihood cluster properties in the spherical and triaxial cases correspond to very different proper motions, when we calculate the likelihood from Equation \ref{eq:LogL}: ($\mu_{\delta}$, $\mu_{\alpha}\mathrm{cos}(\delta)$) = (-2.35, -2.35) and (-3.7,- 5.0) mas yr$^{-1}$, respectively. These give transverse velocities of $v_{tan}$ = 123 km/s (spherical) and $v_{tan}$ = 449 km/s (triaxial) in the Galactic rest frame. Moreover, the LL for the most likely proper motions in the spherical case (LL = -45) is much more strongly peaked with a value considerably higher than in the triaxial case (LL = -82). 

We visualize these results with $N$-body simulations evolving along the most likely orbit in the spherical and triaxial LM10 potentials shown in the right column of Figure \ref{fig:posonly}. The $N$-body particles are over-plotted on the density contours of color-selected Pal\,5 member stars from SDSS (blue). It is evident that the model stream in the LM10 potential does not fit the data well (bottom right). For this particular model the cluster is moving very fast, $v_{tan}$ = 449 km/s, and is on a highly eccentric orbit. It has recently been tidally shocked and has lost a substantial amount of mass, which can be seen as a dense cloud surrounding the cluster. It clearly does not follow the observed ``S''-shape (see zoom in of cluster center), nor the overall curvy morphology of the tails. Instead, the best fit model appears more like a straight line through the data points, which explains why the LL is much lower.

Figure \ref{fig:radvel} shows a comparison of the simulated model streams with the line-of-sight velocities (\citealt{oden09}). The left panel shows the spherical model points, which fit the observed velocities very well even though the proper motion was chosen to match morphology alone. However, the same is not true for the triaxial LM10 case where there is a much stronger velocity gradient in opposite sense to that observed. 

Motivated by the discrepancy in velocities in the triaxial case, we repeat the experiment of finding the most likely configuration of proper motions while now also comparing the \textit{streakline} model streams to observed radial velocities from \citet{oden09} (Equation \ref{eq:LogLtot}). In the spherical case the most likely \textit{streakline} model yields the same values for the two proper motion components as found in Figure \ref{fig:posonly}. 

The left panel of Figure \ref{fig:posvr} shows the results of the parameter space search for the triaxial case. A large discrepancy is still found between the values of the LL in the spherical case (LL = -124) and triaxial case (LL= -180) when we include the radial velocities to our assessment of the likelihood (Equation \ref{eq:LogLtot}). The right panel of Figure \ref{fig:posvr} shows that an $N$-body simulation of the most likely configuration of proper motions in the LM10 potential (where ($\mu_{\delta}$, $\mu_{\alpha}\mathrm{cos}(\delta)$) = (-2.15, -2.4)  mas yr$^{-1}$) yields a significantly better fit to the gradient of the line-of-sight velocities and $v_{tan}$ = 116 km/s. However, the middle panel of Figure \ref{fig:posvr} demonstrates that the morphology of Pal\,5's tidal tails is a poor match to the density contours of SDSS. This explains the difference in the LL between the spherical and triaxial case: the stream appears to ``fan'' out as the debris moves away from Pal\,5. Moreover, this \textit{stream-fanning} explains why this particular proper motion configuration was strongly disfavored when considering morphology alone. 

In summary, in the spherical potential, model streams can easily be found that match the morphology of the Pal\,5 stream and these coincidentally fit the observed line-of-sight velocities. In contrast, the best fit model streams to the morphology in the triaxial potential are much poorer, have a higher transverse velocity (in the following referred to as ``\textit{high-velocity}'' models), and are inconsistent with observed line-of-sight velocities. Model streams in the triaxial potential that match the line-of-sight velocities have similar proper motions to the spherical case, but have \textit{fanned} density distributions that are inconsistent with observations (referred to as \textit{stream-fanning} models). 

However, it is important to note that even if we didn't have the 17 radial velocities along the Pal\,5 stream from \citet{oden09}, we would still have concluded (from the large discrepancy in LL) that the much simpler spherical halo yields better fits to the SDSS data.

\section{Discussion}
\subsection{Exploring the parameter space further}\label{sec:discussion}
In this Section we test whether any unexplored dimensions of parameter space could change the results of Section \ref{sec:results}:
\begin{enumerate}
\item \textbf{Surface density}: we have assumed that all parts of our model streams would be observable, whereas in reality the observations could be showing us only the densest parts of the stream. For example, the \textit{fanned} parts of the streams might not be observable in the SDSS density map due to foreground/background contamination.
\item \textbf{Integration time}: in some cases the extent of our model streams do not span over the full extent of the observations. For example, if the \textit{fanned} parts of the streams are excluded in our model streams, the densest parts could look like thin streams, however for our original choice of a 6 Gyr integration time the model streams are too short to trace the extent of the observations. A longer integration time might lead to longer model streams. 
\item \textbf{Cluster distance}: as we observe the streams in projection, the apparent curvature might change if Pal\,5's present-day distance was varied. For example, the preference of the ``\textit{high-velocity}'' models in the triaxial case leads to a lack of curvature in these model streams due to the proximity to perigalacticon of the stream at this point in phase space. 
\end{enumerate}
To address these three concerns we ran one additional grid of \textit{streakline} models in the triaxial LM10 potential, where we explored these three dimensions in addition to allowing proper motion to vary: (1) we applied density cuts based on the surface density around each \textit{streakline} model point, calculated using a Gaussian density kernel with a width of 1 degree, where we included 100\%, 75\%, 50\% and 25\% of the most dense regions of the streams, (2) we ran the \textit{streakline} models for seven different integration times (4, 5, 6, 7, 8, 9 and 10 Gyr), and (3) we tested various distances (22.6, 23.1, 23.6, 24.1 and 24.6 kpc) to Pal\,5 based on the observational uncertainties (\citealt{dotter11}; Dotter private communication). 

\begin{figure*}
\centerline{\includegraphics[width=1.17\textwidth]{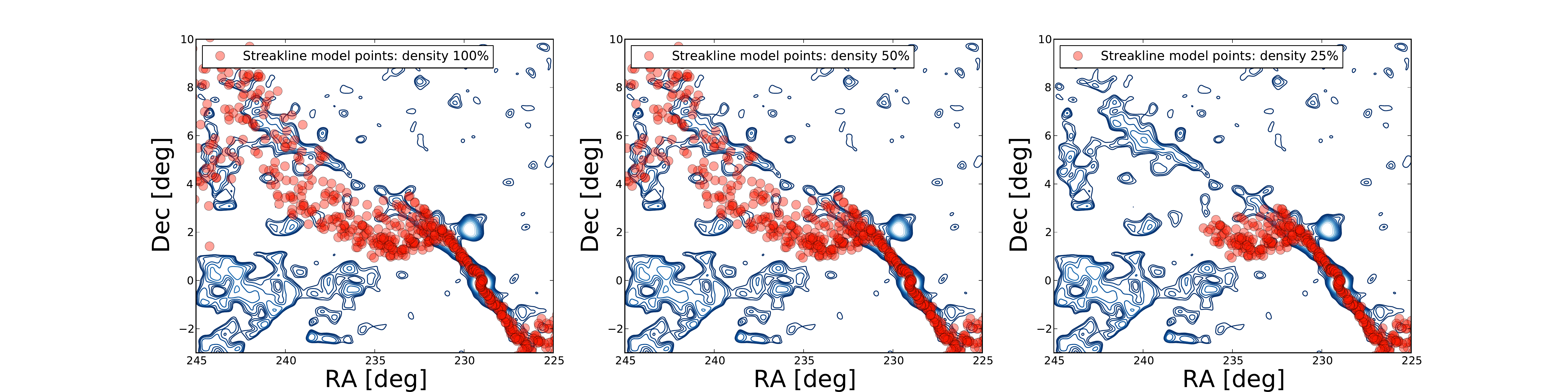}}
\caption{\textit{Streakline} model points (orange) for three different surface density cuts (100\%, 50\% and 25\%) for the most likely \textit{stream-fanning} model over-plotted on SDSS density contours (blue). The most likely \textit{stream-fanning} model does not yield a long, thin stream that fits the SDSS density map.}
\label{fig:StrFan}
\end{figure*}

For all combinations of these parameters, the highest likelihood proper motion configuration was close to the ``\textit{high-velocity}'' models. However, the LL of the secondary, \textit{stream-fanning} peak was more prominent when density cuts were applied. Figure \ref{fig:StrFan} illustrates the success and limitation of the \textit{stream-fanning} model by plotting the most likely model configuration (($\mu_{\delta}$, $\mu_{\alpha}\mathrm{cos}(\delta)$) =  (-2.25,-2.5) mas yr$^{-1}$, 50\% density cut, $t$ = 9 Gyr, $d$ = 23.1 kpc, LL = -83) from the newly explored grid. It is evident that the dense parts of the streams provide a better fit to the SDSS density map. However, this configuration of proper motions does not yield a long, thin stream that could fit the extent of the observational coverage from the SDSS density map, even when the \textit{fanned} parts of the streams are excluded. Hence, the \textit{stream-fanning} peak model still fails to produce a stream that matches the observational data.

\begin{figure}
\includegraphics[width=\columnwidth]{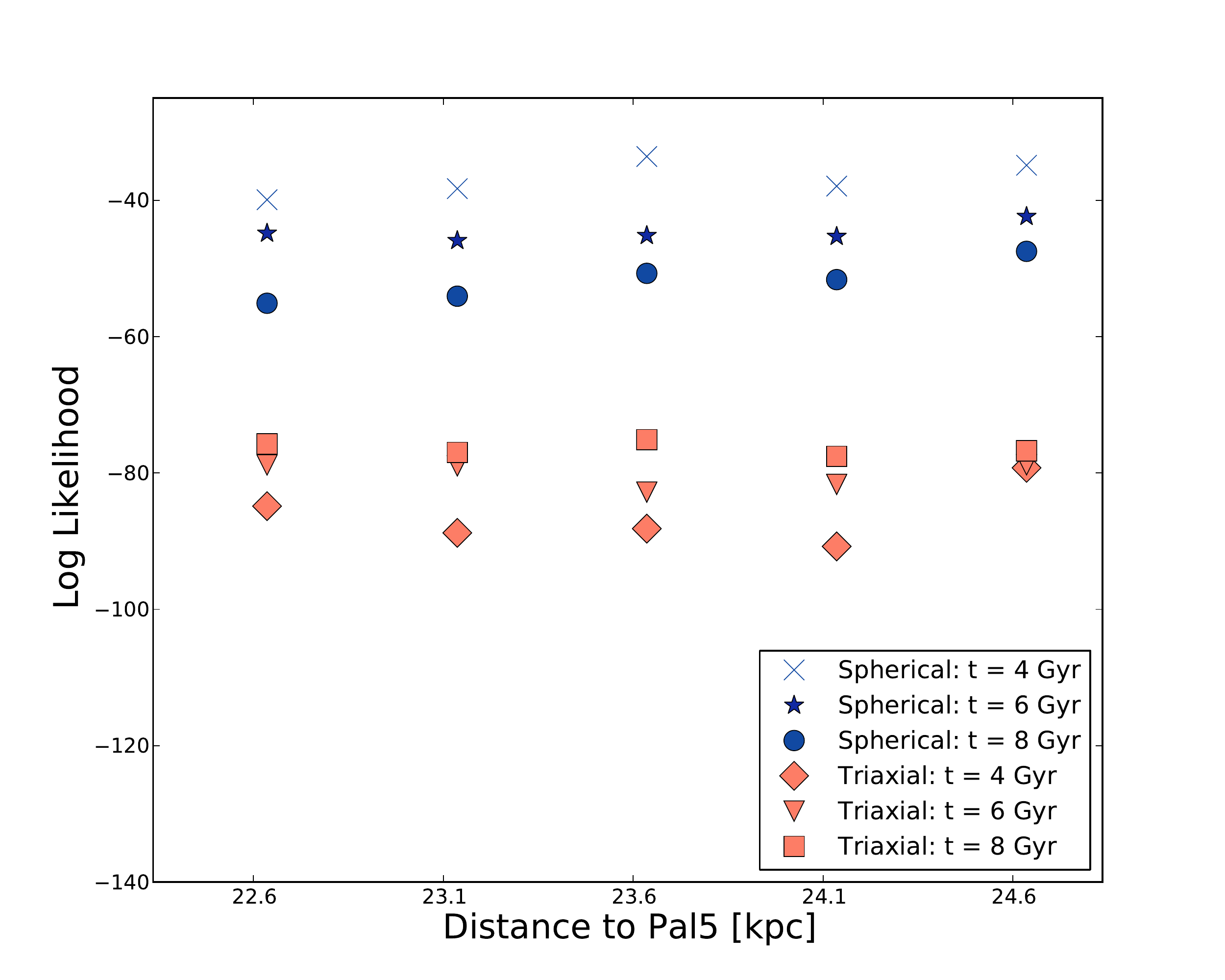}
\caption{Log-likelihoods (Equation \ref{eq:LogL}) of the most likely proper motion configurations for the spherical potential (blue) and triaxial LM10 potential (red) for various distances and integration times. The spherical potential yields much higher likelihoods in all cases.}
\label{fig:LogL}
\end{figure}

Figure \ref{fig:LogL} summarizes the remaining exploration over integration time and various distances. The LL of the most likely \textit{streakline} model is plotted as a function of distance for both the spherical potential and LM10 potential for various integration times. There are no significant changes in the values of the likelihoods with distance. Moreover, it is clear that the stream models in the spherical potential yield much higher values for the likelihoods. 

Other parameters that remain unexplored are the assumed density profile of our cluster (we used a smoothed point-mass for the cluster particle in the \textit{streakline} models) and exploring different velocity properties of the ejected stars. As we are mainly concerned about the morphology of the streams on large scales, where the stars' motions are dominated by the gravitational potential of the galaxy, and since we are mainly interested in the coldest part of the stream, which are given through the over-densities and produced by low-velocity escapers, we do not explore these parameters here. 

Conclusively, these new explorations confirm our findings from Section \ref{sec:results}: the morphology of Pal\,5's tidal tails cannot be reproduced in the triaxial LM10 potential.  

\subsection{Discussion of \textit{stream-fanning}}\label{sec:morph}
As we have shown, it is not possible to produce a thin and curved stream like Pal\,5 in a LM10 potential configuration. The streams are either too straight as the cluster is moving at a very high velocity, or, as the cluster velocity is reduced and the stream gets more curvy, it starts to \textit{fan} out. In this Section, we discuss some tests we have done to provide more information about the process.

We first check the eccentricity of the orbit. In the triaxial case, Pal\,5 could simply be on a significantly more eccentric orbit, which could make its stream appear more as a cloud than a stream (see e.g.~\citealt{johnston08}). However, it is important to note that the orbital parameters such as eccentricity ($e$) and apocenter distance ($R_{apo}$) of the \textit{stream-fanning} orbit shown in Figure \ref{fig:posvr} ($e$ = 0.34, $R_{apo}$ = 18.6 kpc) are very similar to the parameters for the spherical model stream's orbit shown in the top panel of Figure \ref{fig:posonly} ($e$ = 0.39, $R_{apo}$ = 19.4 kpc). Here eccentricity is defined as: $e = \frac{R_{apo}-R_{peri}}{R_{apo}+R_{peri}}$.

We next investigate the orbital classes of the \textit{stream-fanning} model orbits to see whether these orbits are loop or box orbits. Regular loop orbits preserve a sense of rotation about the long or short axis of the potential, whereas box orbits may approach close to the center of the potential and have (on average) smaller pericenters, both of which would affect the disruption of the cluster and subsequent evolution of the debris.

To explore the orbits around the  \textit{stream-fanning} region in LM10, we investigate a grid from $\mu_{\delta}$ $-2.0$ to $-2.3$\,mas\,yr$^{-1}$ and $\mu_{\alpha}\mathrm{cos}(\delta)$: $-2.3$ to $-2.55$\,mas\,yr$^{-1}$ (see left panel of Figure \ref{fig:posvr}) with step sizes of 0.05 mas yr$^{-1}$, while fixing $v_r$ = -58.7 km/s, $d$ = 23.6 kpc and $t$ = 6 Gyr. The streams from these initial conditions were all \textit{fanned}. These orbits have $v_{tan}$ = 100 -- 137 km/s. To determine the orbital class of each orbit from the investigated grid, we check for a sign change in the angular momentum about any of the coordinate axes. We find that all orbits in the \textit{stream-fanning} region preserve the sign of their angular momenta which suggests they are on loop orbits. An additional test was made by integrating all the final cluster particle positions in the \textit{fanned} tails from Figure \ref{fig:posvr} backwards in LM10 for 6 Gyr.  All of these particles were also on loop orbits.

Another possibility is that the debris is \textit{fanned} in LM10 due to Pal\,5 and its debris being on chaotic orbits. That is, if Pal\,5 was on a chaotic orbit, the orbits of debris stars could diverge significantly from the cluster's trajectory (an example of this was shown in Figure 11 of \citet{fardal14}). To check if the \textit{stream-fanning} of Pal 5 in LM10 is due to chaos, we measure the Lyapunov spectrum \citep[e.g.][]{Froeschle97, skokos10} of the cluster orbit for each orbit in the \textit{stream-fanning} grid. We use the maximum Lyapunov exponent, $\lambda_{\rm max}$, for an orbit to ask whether it is chaotic over a timescale comparable to the interaction time of the cluster with the host potential. We do this by measuring the Lyapunov time:
\begin{equation}
t_{\rm lyap} = \frac{1}{\lambda_\mathrm{max}}
\label{eq:lyapt}
\end{equation}
We find that some orbits are stochastic, but have Lyapunov times $\gtrsim 20~\mathrm{Gyr}$ (e.g., many times the integration time of any of the stream models above). 

Finally, we attempt to compute the Hessian of the Hamiltonian for this potential in action-space, evaluated at the orbit of the cluster. For a thin stream to form, the Hessian should be dominated by a single eigenvalue \citep[see, e.g., ][]{bovy14, sanders13}, however since the fanned debris is instead spread over two dimensions, we aim to check whether the Hessian along Pal 5's orbit is instead dominated by two eigenvalues of comparable magnitude.  To convert from phase-space to action-angle coordinates, we use the algorithm and code from \citet{sanders14}. We find that for and around the stream-fanning orbit of Pal 5, action-angle solutions cannot be found. This could be an indication of chaos, but the long Lyapunov times indicate that perhaps a different mechanism is at play. When investigating the time variability of the ``toy actions'' --- which, in the method of \citet{sanders14} are modeled as a Fourier series expansion in a combination of the target actions, generating function, and toy angles --- we find that there are many prominent ``spikes'' that correspond to the orbit passing through the midplane of the disk potential. The transformation to action-angle coordinates is failing due to the non-adiabatic forcing of the disk potential, however it is unclear whether this is a failure of the method or because the disk potential is destabilizing the orbit itself and making it more irregular (where actions do not exist). Thus, we are unable to compute the Hessian.

From our preliminary investigation, we conclude that the Pal 5 stream in LM10 does not appear to be \textit{fanned} due to Pal 5 being on an orbit that takes it closer to the Galactic center either due to high eccentricity or box orbits, nor due to Pal 5 being on a chaotic orbit in LM10. We additionally note that we have tried and failed to use the machinery of \citet{sanders14} to compute actions and angles for the stream-fanning orbit. Thus, the origin of \textit{stream-fanning} remains uncertain. We are currently studying this phenomenon and postpone a more thorough exploration of its origin to our forthcoming work (Price-Whelan et al., in prep).

\section{Conclusion} \label{sec:conclusion}
In this paper we used the thin and curved morphology of the Pal\,5 stream alone (without the need for additional dimensions of information) to rule out the triaxial shape of the Galactic halo potential as suggested by \citet{lawmajew10}. That the Galactic halo potential is not of the form suggested by \citet{lawmajew10} is not very surprising or new, as the LM10 potential has been pointed out to have several issues (see Section \ref{sec:intro}).
However, we found that models evolved along Pal\,5-like orbits in this particular triaxial potential generally exhibited an unusual morphological signature - which we dubbed \textit{stream-fanning}. 

There are several examples already in the literature where morphology alone has been used to rule out certain forms of the Galactic potential. For example: the degree of alignment of the tails from Sagittarius along a single great circle was used to discuss how far from spherical the Galaxy's potential might be \citep{ibata01,johnston05,fell06}; \citet{lux12} showed that the path of the 45 degree long stream associated with the globular cluster NGC 5466 was incompatible with spherical or prolate halo models of a variety of parametric forms; and (as mentioned in the introduction) the precession angle between successive apocenters traced by Sagittarius' debris has been interpreted as an indicator of the radial density profile of the dark matter halo \citep{belokurov14,gib14}.

The discovery of \textit{stream-fanning} as a phenomenon sensitive to the triaxiality of the mass distribution adds a new approach to this toolkit of potential measures.
We already know of many other thin streams at different radii and orbiting with different orientations throughout the Milky Way that could be investigated using this new approach (e.g., NGC 5466 (\citealt{grilljon06}), GD-1 (\citealt{grill06}), Orphan (\citealt{belokurov07}, \citealt{casey13}), Acheron, Cocytos, Lethe, Styx (\citealt{grill09}), Triagulum (\citealt{bonaca12}), Ophiuchus (\citealt{bernard14})), and fainter streams are likely to be discovered in the future. Mapping these even more distant structures in velocity can only be more challenging.
While the origin of \textit{stream-fanning} is still under investigation (Price-Whelan et al., in prep), this first study of Pal 5 indicates the promise of using the {\it absence} of stream fanning in observed streams as a means to rule out classes of potentials.
Collectively, the existence and location of these thin streams should provide broad but powerful constraints on the shape of the MW potential on large scales.


\acknowledgements
We thank the referee for a very detailed report, which helped to improve the quality of the paper significantly. KVJ thanks Hanni Lux for the initial collaborative work that led to this project. This work was supported in part by the National Science Foundation under Grant No. AST-1312196. AHWK acknowledges support from NASA through Hubble Fellowship grant HST-HF-51323.01-A awarded by the Space Telescope Institute, which is operated by the Association of Universities for Research in Astronomy, Inc., for NASA, under contract NAS 5-26555. APW is supported by a National Science Foundation Graduate Research Fellowship under Grant No. 11-44155. This research made use of Astropy, a community-developed core \texttt{Python} package for Astronomy \citep{astropy13}.


\end{document}